\documentclass[preprint,aps,showpacs]{revtex4}

\usepackage{amsmath}
\usepackage{graphicx}
\usepackage{dcolumn}

\begin{document}


\title{Quantum interference of multimode two-photon pairs with a Michelson
interferometer}


\author{Fu-Yuan Wang}
\author{Bao-Sen Shi}
\email{drshi@ustc.edu.cn}
\author{Guang-Can Guo}
\affiliation{Key Laboratory of Quantum Information,University of Science and Technology of China,Hefei 230026,China 
}


\date{\today}

\begin{abstract}
We experimentally observe the two-photon interference of multimode photon
pairs produced by an optical parametric oscillator far below threshold via a
michelson interferometer, which shows a multipeaked structure. We find that the correlation function when the
interferometer is 
unbalanced is clearly dependent on the path difference and phase between
two interfering beams, but the shape of correlation function in balanced case
is independent on the small path difference and phase beside the height. All
experimental results are well agreed with the theoretical prediction.
\end{abstract}


\maketitle

\begin{figure}[t]
  \begin{center}
	\includegraphics[width=0.5\textwidth]{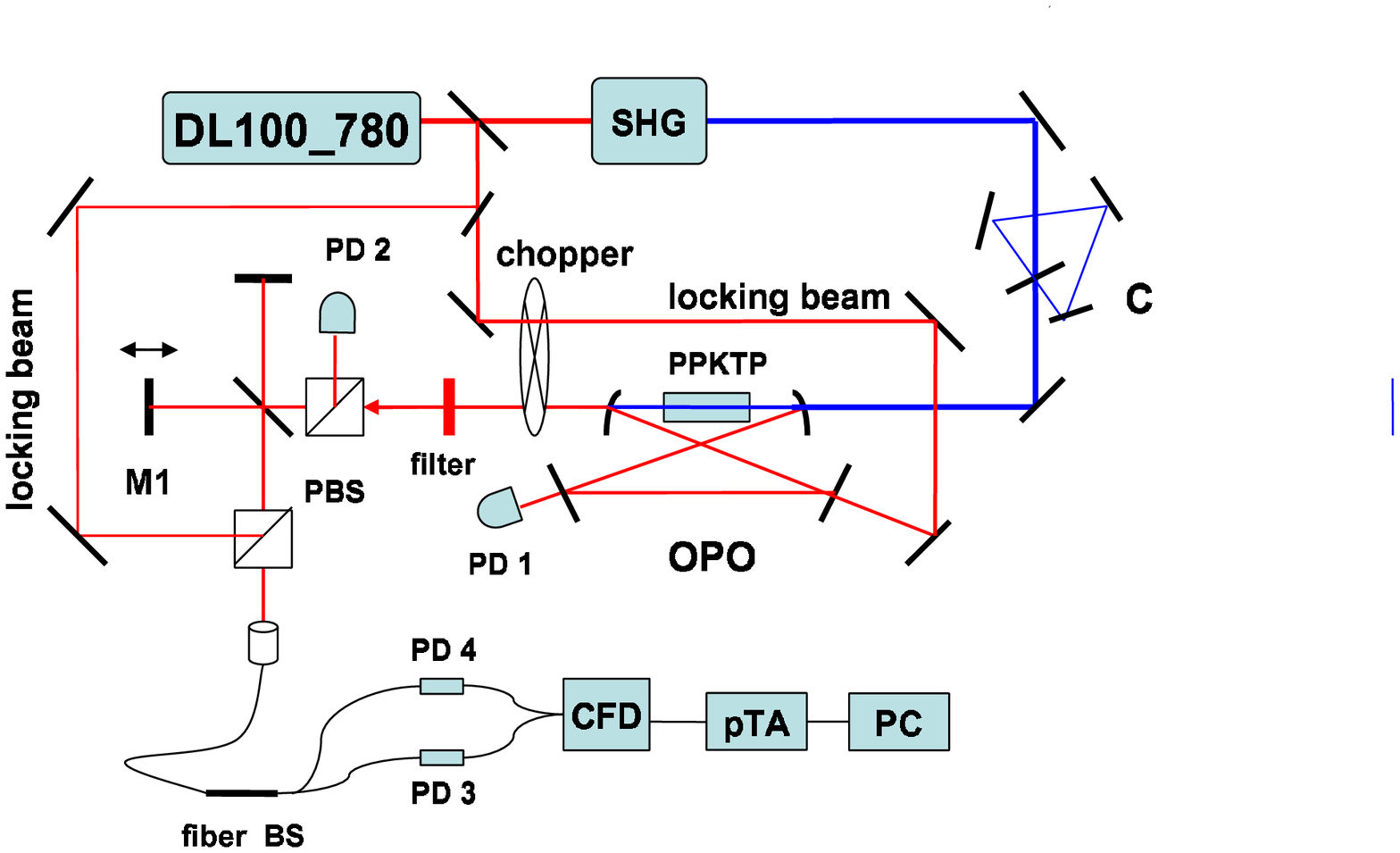}
  \end{center}
  \caption{Schematic of the experimental setup. DL100\_780, diode laser; PBS,
  polarization beam splitter; PD1 and PD2, photodetectors for locking; PD3 and
  PD4, avalanche photodetectors; fiber BS, fiber beam splitter; C, triangle
  cavity; pTA, picosecond time analyzer}
  \label{fig:setup}
\end{figure}

\indent Entangled photon pair source, as an essential tool, plays an important
role in quantum information processing and quantum optics field\cite{PhysRevLett.67.661,PhysRevLett.70.1895,PhysRevLett.76.4656,Bouwmeester575,deutsch1992}. By far, the
common way to generate an entangled photon pair is the process of spontaneous
parametric down-conversion(SPDC) in a nonlinear crystal\cite{PhysRevLett.25.84}, for it is most accessible and controllable
in present technique. One drawback of the photon generated via SPDC is its
wide spectrum, which makes its interaction with atoms very difficult. In order
to solve this problem, several groups generate a narrow-band photon pair via
an optical parametric oscillator(OPO) far below
threshold\cite{PhysRevLett.83.2556,PhysRevA.68.015803,scholz:191104,kuklewicz:223601,PhysRevA.69.035801,wangmultimode}. The property of the
two photons produced by this way enable us to directly observe their
correlation function by coincident counting, and an interferometer can be used
to realize this goal. 
Goto \textit{et.al.} recently reported the two-photon interference of
multimode two-photon pairs with an unbalanced Mach-Zehnder
interferometer\cite{PhysRevA.69.035801}. The time correlation between the
multimode two photons has a multipeaked structure. In
their experiment, the propagation time difference T between the short and long
paths in the interferometer is $\tau_r/2$, where $\tau_r$ is the round-trip time of the
OPO cavity. Their results show that the property of the multimode two-photon
state induces two-photon interference depending on the delay. In this paper,
we report on an observation of quantum interference of multimode two-photon
pairs generated via an OPO far below threshold with a Michelson interferometer.
In the experiment, we not only discuss the case in which the interferometer is
 in highly unbalanced, but also consider the case when the interferometer is
 balanced. In highly unbalanced case, the path propagation difference between
 the short and long paths is nearly equal to $\tau_r/3$. In this situation,
 only two-photon interference occurs, there is no single photon interference
 because of quite short coherence length($<$100 $\mu$m) measured in our
 experiment\cite{shicolength}. The time
 correlation between the multimode two photons still has a multipeaked
 structure. We find that the shape of correlation fringe is different from
 that reported in Ref.\cite{PhysRevA.69.035801}. Therefore we conclude that
 the shape of correlation is dependent on the path difference between two
 interfering beams. In balanced case, we consider two different situations:
 one is that the interferometer is perfectly balanced, which means that the
path difference is almost zero. In this situation, both two-photon
interference and single- photon interference exist simultaneously. The time
correlation has multipeaked structure, but is very different from that in
highly unbalanced case. The shape of the fringe is independent on the phase
between two interfering beams beside height. We also consider another
situation: the interferometer is almost balanced, the path difference is
slightly larger than the coherent length of the single photon. In this 
situation, there is only two-photon interference. The time correlation
observed still has the multipeaked structure, is similar to that in perfectly
balanced situation, and the shape of fringe is also independent on the phase
between two interfering beams beside the height. All experimental results are
well agreed with the theoretical prediction. 
\\
\indent A schematic drawing of the experimental setup is shown in Fig.
\ref{fig:setup}. A cw grating-stabilized external diode laser (Toptical DL100)
of wavelength 780 nm is used to generate UV light at 390 nm via a frequency
douber, which consists of a symmetric bow-tie cavity with a 10 mm long type-I
phase-matched periodically poled KTiOPO$_4$(PPKTP) inside. The  frequency of
the laser is precisely locked to Rb atom transition frequency using the
saturated absorption technique. The frequency of the doubling cavity is locked
to frequency by PDH method\cite{PDHmethod}. About 50 $\mu$W UV light at 390 nm is input to an
OPO far below threshold, which consists of a 10 mm long type-I phase-matched
PPKTP and a symmetric bow-tie cavity. The triangle cavity C is used to get
mode-matched between SHG cavity and OPO cavity. A chopper is used to cut the
photons of locking light reflected from the surface of the crystal to avoid
possible background noise. The photon generated is multimode photon because of
no mode-selected cavity used, and has the comb-like shape of the
spectrum\cite{shicolength}. The
outputs from OPO are input into a Michelson interferometer. A red filter id
used to cut the remaining UV light. The path difference between two
interfering beams can be adjusted by moving the mirror M1, which is mounted
a piezoelectric transducer(PZT). Both PZT and mirror are fixed on a
translation stage. The relative phase between two interfering beams can be
actively controlled. The one output of the interferometer is connected to a
50/50 fiber beam splitter(NEWPORT P22s780BB50). Each out port of the fiber BS
is connected to an avalanche photon detector(PerkinElmer SPCM-AQR-14-FC). The
outputs from detectors are sent to a coincidence circuit for coincidence
counting which mainly consists of a picosecond time analyzer(ORTEC, pTA9308)
and a computer.
\\
\indent First, we show our experimental result in unbalanced case. In this
case, the path difference between two interfering beams $\Delta L$ is
about 170 mm. Before discussing the experimental results, we give the theory about the time 
correlation function in the unbalanced case which mainly comes from
Ref.\cite{PhysRevA.69.035801} . The output operator $a(t)$ of the interferometer can be expressed by
\begin{align}
  a(t)=\frac{a_{out}(t-T_S)+ia_{vac}(t-T_S)}{2}+\frac{a_{out}(t-T_L)-ia_{vac}(t-T_L)}{2}
\end{align}
Here T$_L$, T$_S$ are the propagation time of the photons along the short and
long paths, respectively. $a_{vac}(t)$ is an annihilation operator of the vacuum entering the
interferometer. $a_{out}(t)$ is the output operator of the OPO which can be
defined as
\begin{align}
  a(t)=\frac{1}{\sqrt{2\pi}}\int d\omega a(\omega)e^{-i(\omega_0+\omega)}
\end{align}
Where $\omega_0$ is the degenerate frequency of the OPO. The time correlation function is defined as follows
\begin{align}
  \Gamma^{(2)}(\tau)=\langle
  a^+(t)a^+(t+\tau)a(t+\tau)a(t)\rangle\label{corfunc}
\end{align}
And after dropping the single photon interference terms and considering the
probability distribution of timing jitter of detectors, Eq.(\ref{corfunc}) is
expressed as
\begin{align}
  \Gamma_{unbal}^{(2)}(\tau)=C_1\left[4\Gamma_{ave}^{(2)}(\tau)\cos^2\theta+\Gamma_{ave}^{(2)}(\tau-T)+\Gamma_{ave}^{(2)}(\tau+T)\right]+C_2\label{expeq}
\end{align}
with
\begin{align}
  \Gamma_{ave}^{(2)}(\tau)=&
  e^{-\Delta\omega_{opo}|\tau-\tau_0|}\sum_n\left(1+\frac{2|\tau-n\tau_r-\tau_0|\ln2}{T_D}\right)\nonumber\\
  &\times\exp\left(-\frac{2|\tau-n\tau_r-\tau_0|\ln2}{T_D}\right)
\end{align}
\begin{figure}[t]
  \begin{center}
	\includegraphics[width=0.5\textwidth]{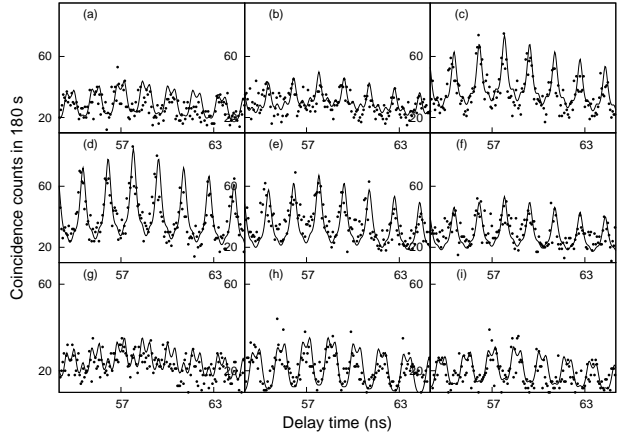}
  \end{center}
  \caption{Experimental results. The phase $\theta$ increases stepwise by
  $\arccos$ j/4 from (a) to (i) (j=-4,-3,$\cdots$,4). The dots represent the
  measured data. The lines are fits by Eq. (\ref{expeq}).}
  \label{fig:unbal}
\end{figure}
Here C$_1$ is a constant; C$_2$$\propto$$|\epsilon|^2\Gamma^{(2)}(0)/\Delta\omega_{opo}^2$, $\epsilon$ is
single-pass parametric amplitude gain; $\Delta\omega_{opo}$ is the bandwidth
of the OPO; T$_D$ is the resolving time of the detectors which is about 220 ps in our experiment; $\tau_0$ is an
electric delay, T=T$_L$-T$_S$. We measure the coincidence counts at
$\cos\theta$=$j/4\;(j=-4,-3,\cdots,4)$, where $\theta$ is the phase difference
between the two arms of the interferometer. 
 The experimental
results are shown in Fig. \ref{fig:unbal}. We set $\Delta L$ as 170 mm 
which is about $\tau_rc/3$ long. The constant parameters are set as follows: $\tau_0$=55 ns, 
$\tau_r$=1.63 ns and $\Delta\omega/2\pi$=7.8 MHz. The time correlation has a multipeaked structure, the shape of the fringe is different from that shown in Ref. \cite{PhysRevA.69.035801}, this concludes that time correlation is dependent on the path difference. In Fig. 
\ref{fig:unbal}, the deviation of the points from the fitted lines is probably 
due to the small $\tau_r$ which determines the distance between peaks. If
$\tau_r$ is larger, every peak could be more distinguished from each other
under the condition of the big resolving time of the detectors, which could achieve better 
visibility. The fluctuation of phase difference may lead to the deviation as 
well.  Fig. \ref{fig:phase} shows the deviation between the locking points of the interference and 
the fitting points. We find that points (g) and (h) have a little large 
deviation. This may also come from the fluctuation of phase difference and
misalignment of the interferometer, etc.
\begin{figure}[t]
  \begin{center}
	\includegraphics[width=0.5\textwidth]{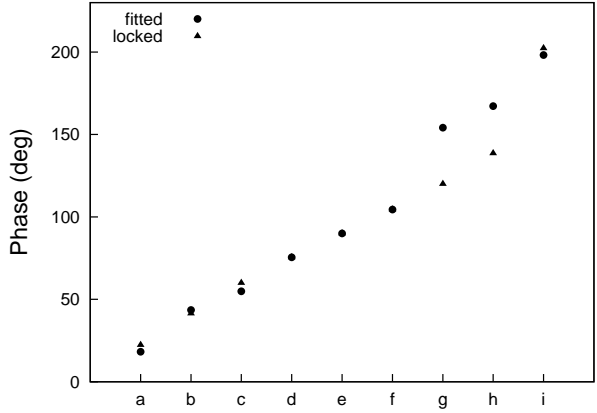}
  \end{center}
  \caption{ Phase (fitting) determined by the fitting is plotted against phase
  (locking) which is the phase locked experimentally. Triangles represent the
  experimental date and points represent the fitting one. The letter at
  horizontal axis(a,b,$\cdots$,i) represents the picture of the same letter in
  Fig. \ref{fig:unbal}}
  \label{fig:phase}
\end{figure} 
\\
\indent Next, we discuss the case in which the two interfering beams in the
interferometer are almost balanced. According to our measurement on the
coherence length of the single photon , which is about 90
$\mu$m\cite{shicolength}, we divide
this case into two different situations: in situation 1, two interfering beams are
perfectly balanced, which means there is a single-photon interference besides
two-photon interference in the interferometer. In this situation, the time
correlation function can be expressed as follow
\begin{align}
  \Gamma_{a}^{(2)}=&\frac{1}{4}\left[\Gamma_{ave}^{(2)}(\tau)(1+\delta)+\delta\Gamma^{(2)}(0)\right](\cos\theta+1)^2\nonumber\\
  =&\frac{1}{4}\left[C_1\Gamma_{ave}^{(2)}(\tau)+C_2\right](\cos\theta+1)^2\nonumber\\
  \simeq&\frac{1}{4}\Gamma_{ave}^{(2)}(\tau)(\cos\theta+1)^2\qquad(\delta\ll1)
\end{align}
In situation 2, two interfering beam are roughly balanced, which means the path
difference between two interfering beams is slightly larger than coherence
length of the single photon. Therefore, there is no single-photon interference
besides two-photon interference. In this situation, the time correlation function
is follow
\begin{align}
  \Gamma_{b}^{(2)}=&\frac{1}{4}\Gamma_{ave}^{(2)}(\cos^2\theta+\frac{1+3\delta}{2})+\frac{1}{4}\delta\Gamma^{(2)}(0)\nonumber\\
  =&\frac{1}{4}\Gamma_{ave}^{(2)}(\tau)(\cos^2\theta+C_1')+C_2'\nonumber\\
  \simeq&\frac{1}{4}\Gamma_{ave}^{(2)}(\tau)(\frac{1}{2}\cos2\theta+1)\qquad(\delta\ll1)
\end{align}
where C$_1$, C$_2$, C$_1'$ and C$_2'$ are constants;
$\delta$=$4|\epsilon|^2/\Delta\omega_{opo}^2$.  The period of $\Gamma_{b}^{(2)}$ is as twice as that of
$\Gamma_{a}^{(2)}$.
\\
\begin{figure}[t]
  \begin{center}
	\includegraphics[width=0.5\textwidth]{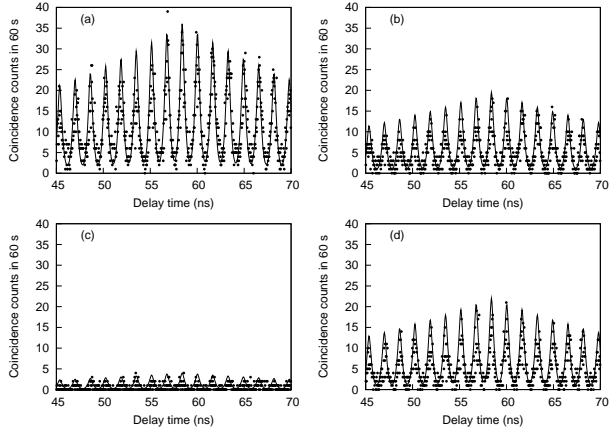}
  \end{center}
  \caption{The time correlation functions of two-photon interference at
  balanced case are measured at $\theta$=0, $\pi/2$, $\pi$ from (a) to (c).
  (d) is the function at two-photon interference at an arbitrary phase with
  $\Delta L$=0.74 mm for
  compare.}
  \label{fig:timebal}
\end{figure}

\begin{figure}[h]
  \begin{center}
	\includegraphics[width=0.5\textwidth]{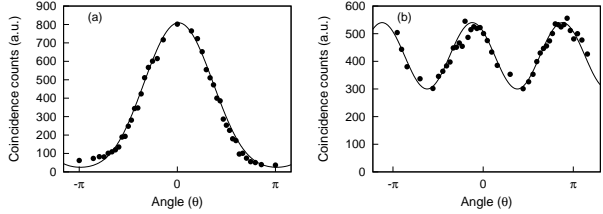}
  \end{center}
  \caption{(a) and (b) are the coincidence counts of   two-photon
  interference in $\Delta L$$\approx$0 and $\Delta L$=0.74 mm, respectively.}
  \label{fig:ccbal}
\end{figure}
\indent The experiment results are shown in Fig. \ref{fig:timebal} and Fig.
\ref{fig:ccbal}. Fig. \ref{fig:timebal}(a)-\ref{fig:timebal}(c) indicate the
time correlation functions at $\theta$=0, $\pi/2$ and $\pi$ with $\Delta
L$$\simeq$0, respectively. The
shapes of three pictures are almost same except the height. The height of the figure in Fig. \ref{fig:timebal}(c) is not exactly
zero, because of the phase fluctuation as well as the imperfect 50/50 BS in
the
interferometer. The experimental results clearly show that time correlation still has the multipeaked structure, but it is very different from that in unbalanced situation. Fig.
\ref{fig:timebal}(d) shows the result when $\Delta L$=0.74 mm. From the
expression of $\Gamma_b^{(2)}$, we know the shape of time correlation
function is not dependent on phase. We measure it at an arbitrary
phase and find it is true. Therefore, we conclude that the shape of the correlation
function is independent on the phase and small difference between two
interfering beams.
\\
\indent In addition, we measure the coincident counts against the phase
between two interfering beams. Fig. \ref{fig:ccbal}(a) and Fig.
\ref{fig:ccbal}(b) shows the results in situation 1 and 2, respectively.  The visibility of the Fig.
\ref{fig:ccbal}(a) is about 88\% against the perfect case of 100\% and that
of Fig. \ref{fig:ccbal}(b) is about 29\% against the perfect case of 50\%. The
possible reason are follow: the shifting of light from ECDL during the experiment; the small shift of
mirror positions; the phase fluctuation and so on. Comparing Fig.
\ref{fig:ccbal}(a) and \ref{fig:ccbal}(b), we can see clearly that the period
of oscillator in situation 2 is as twice as that in situation 1.
\\
\indent In summary, in this  work we observe the quantum interference
of multimode two-photon pairs produced by an OPO with an Michelson
interferometer which shows a multipeaked structure. We find that the correlation function is dependent on the path
difference and phase between two interfering beams in unbalanced case, which
is agree with the Ref. \cite{PhysRevA.69.035801}. Furthermore, we find that
the time correlation shape in balanced case  is independent on the small path difference and
phase, beside the height, compared with that in unbalanced case. All
experimental results are well agreed with theoretical prediction.

\begin{acknowledgments}
We thank Mr. J. S. Xu and Dr. C. F. Li for their kindly lending of pTA.
This work is supported by National Natural Foundation of Science, (Grant
No. 10674126), National Fundamental Research Program (Grant No.
2006CB921900), the Innovation fund from CAS, Program for NCET.

\end{acknowledgments}


\end{document}